\documentclass[useAMS,usenatbib]{mn2e}
\usepackage{epsfig,amssymb,aas_macros}
\title[CO in the halo of a HzRG]{CO line emission in the halo of a radio
galaxy at z=2.6\thanks{Based on observations collected at the IRAM Plateau de
Bure Interferometer (PdBI).}}

\author[N. Nesvadba et al.]{ \parbox[h]{\textwidth}{N.~P.~H.~Nesvadba$^{1,2}$\thanks{E-mail:
    nicole.nesvadba@ias.u-psud.fr}, R.~Neri$^{3}$, C.~De Breuck$^{4}$,
    M.~D.~Lehnert$^{2}$, D.~Downes$^{3}$, F.~Walter$^{5}$, A.~Omont$^{6}$, F.~Boulanger$^{1}$,
    N.~Seymour$^{7}$}\\
$^{1}$Institut d'Astrophysique Spatiale, Universit\'e Paris Sud 11, Orsay, France \\
$^{2}$GEPI, Observatoire de Paris, CNRS, Universit\'e Denis Diderot, Meudon, France\\
$^{3}$Institut de Radio Astronomie Millimetrique (IRAM), St. Martin d'Heres, France\\
$^{4}$European Southern Observatory, Karl-Schwarzschild Strasse, Garching
bei M\"unchen, Germany\\
$^{5}$Max Planck Institut f\"ur Astronomie, Heidelberg, Germany\\
$^{6}$Institut d'Astrophysique de Paris, CNRS, Universit\'e Pierre et Marie Curie, Paris \\
$^{7}$Mullard Space Science Laboratory, UCL, Holmbury St. Mary, Dorking,
Surrey, RH5 6NT, UK}

\begin{document}



\maketitle

\label{firstpage}

\begin{abstract}
We report the detection of luminous CO(3--2) line emission in the halo of the
z$=$2.6 radio galaxy (HzRG) TXS0828+193, which has no detected counterpart at
optical to mid-infrared wavelengths implying a stellar mass $\lesssim$
few$\times 10^9$ M$_{\odot}$ and relatively low star-formation rates. With the
IRAM PdBI we find two CO emission line components at the same position
at $\sim$80 kpc distance from the HzRG along the axis of the radio jet,
with different blueshifts of few 100 km s$^{-1}$ relative to the HzRG and a
total luminosity of $\sim$ 2$\times$ $10^{10}$ K km s$^{-1}$ pc$^2$ detected
at a total significance of $\sim$ 8$\sigma$. HzRGs have significant galaxy
overdensities and extended halos of metal-enriched gas often with embedded
clouds or filaments of denser material, and likely trace very massive
dark-matter halos. The CO emission may be associated with a gas-rich, low-mass
satellite galaxy with very little on-going star formation, in contrast to all
previous CO detections of galaxies at similar redshifts.  Alternatively, the
CO may be related to a gas cloud or filament and perhaps jet-induced gas
cooling in the outer halo, somewhat in analogy with extended CO emission found
in low-redshift galaxy clusters.
\end{abstract}


\section{Introduction}
\label{introduction}
The most vigorous starbursts in the Universe occurred in massive galaxies
during the most active phase of galaxy evolution and AGN activity, at
redshifts z$\sim 2-3$. These galaxies formed most of
their stellar mass of a few $\times 10^{10-11}$ M$_{\odot}$ in short bursts of
few $\times$100 Myrs with star-formation rates of several 100 M$_{\odot}$
yr$^{-1}$ \citep[e.g.,][]{smail02,archibald01,reuland04}. Luminous CO emission
observed at millimeter wavelengths is
typically viewed as the most direct sign of the immense reservoirs of cold
molecular gas necessary to fuel these starbursts \citep[e.g.,][]{greve05}.

Powerful high-redshift radio galaxies (HzRGs) may host the most extreme
starbursts at high redshift \citep[e.g.,][]{seymour08}, seen in a short, but
critical phase of their 
evolution dominated by strong AGN feedback \citep{nesvadba06, nesvadba07b,
nesvadba08b}. Bright K-band magnitudes suggest HzRGs are among the most
massive galaxies at all cosmic epochs \citep[e.g.,][]{debreuck02}, a
conclusion recently confirmed through rest-frame near-infrared
photometry. \citet{seymour07} find stellar masses of several $10^{11}$
M$_{\odot}$, factors of a few larger than typical masses of submillimeter
galaxies at similar redshifts \citep[$10^{10.5}$
  M$_{\odot}$,][]{smail04}. Luminous CO emission has been found in several
HzRGs \citep[e.g.,][]{papadopoulos00, debreuck05, klamer05}, and recently also
in a satellite galaxy of a HzRG \citep{ivison08}.

HzRGs reside in particularly rich environments, and are often surrounded by
several 10s of companion galaxies \citep[e.g.,][]{lefevre96,kurk04,venemans07}
as well as extended halos of ionized and neutral gas. Faint, diffuse
Ly$\alpha$ emission extends to radii well beyond the inner halo, where the gas
is strongly disturbed by the radio jet. \citet{villar02,villar03} trace
ionized gas out to radii of $> 100$ kpc with relatively quiescent kinematics
and CIV emission line ratios implying near-solar metallicites out to large
radii.  Deep Ly$\alpha$ absorbtion troughs reveal neutral gas,
likely in clouds or filaments \citep[e.g.,][]{vanojik97}.

Using the IRAM Plateau de Bure Interferometer we detected luminous CO(3--2)
line emission in the halo of the z$=$2.6 HzRG TXS0828$+$193, with a
luminosity of $2\times 10^{10}$ K km s$^{-1}$ pc$^2$. Deep photometry
from the rest-frame UV to mid-infrared including MIPS 24$\mu$m imaging does
not reveal a counterpart within the 5\arcsec\ beam, implying a very small
associated stellar mass and low star-formation rates. These are very unusual
properties for a high-redshift CO emitter, and we discuss possible scenarios
for its nature. Throughout the paper, we adopt a flat H$_0 =$70 km s$^{-1}$
Mpc$^{-3}$ concordance cosmology with $\Omega_{\Lambda} = 0.7$ and $\Omega_{M}
= 0.3$.

\section{Observations and ancillary data} 
\label{sec:observations}
We observed TXS0828+193 with the IRAM Plateau de Bure Interferometer PdBI
\citep{guilloteau92} in the D configuration. At z$=$2.6, CO(3--2) falls at
96.6 GHz and into the 3 mm atmospheric window. We reached an 0.3
mJy/beam rms and a beam size of 5.3\arcsec$\times$4.6\arcsec\ ($42\times 37$ kpc
at z$=2.6$). On-source integration time was 12.9 hrs under normal to good
conditions with 6 antennae and system temperatures $< 150$ K. Data were
calibrated using the CLIC package and with MWC349 as flux calibrator. We
combined both polarizations and rebinned the data to a resolution of 30 km
s$^{-1}$. The PdBI
receivers covered a window of $\sim 2600$ km s$^{-1}$.
\begin{figure} \centering
\epsfig{figure=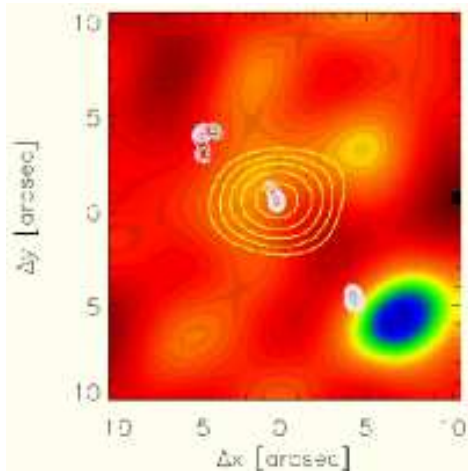,width=0.35\textwidth,angle=0}
\caption{CO(3--2) line image, integrated over the full wavelength range of the
two components. Line emission is not spatially resolved with a
5\arcsec\ beam. Thick yellow and thin white contours show the non-thermal
continuum observed at 3mm with the PdBI and at 8.4 GHz with the VLA,
respectively. \label{fig:txs0828}}
\end{figure}

We also include deep Palomar WIRC K-band imaging, and SPITZER IRAC 3.6$\mu$m and
MIPS 24$\mu$m photometry, as well as archival HST WFPC2
F606W imaging. C. Carilli kindly provided his VLA 8.4 GHz A-array map of
TXS0828+193 first published in \citet{carilli97}. We detected the 
millimeter continuum of the radio core of TXS0828+193. Thus, we could
align the PdBI data with our line-free K band continuum image of TXS0828+193
\citep[][]{nesvadba08b}, assuming that the AGN is at the center of the
galaxy. The K band data then served as a reference to align all other images.

\section{CO emission in the halo of a HzRG}
\label{sec:observation} 
We identify luminous CO(3--2) emission in the halo of TXS0828+193 at a
distance of $\sim 10$\arcsec\ (80 kpc) south-west from the radio galaxy
(Fig.~\ref{fig:txs0828}). The 5\arcsec\ beam corresponds to an upper size
limit of $\sim 40$ kpc. The emission line region is aligned with the axis of
the radio jet but $2.5\arcsec$ (20 kpc) WSW in projection from the radio lobe,
and hence at a larger radius from the central galaxy. Integrating over the
full line we reach $8 \sigma$ significance. To further ensure the robustness
of our measurement, we split the data into two subsamples with consistent
results in each subset.

\begin{figure}
\centering
\epsfig{figure=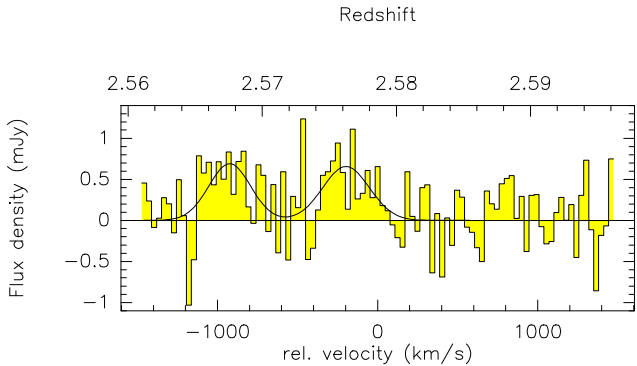,width=0.27\textwidth,angle=270}
\caption{Integrated spectrum of the CO emitter south-west from
TXS082+193. The yellow histogram marks the data, with a Gaussian fit of two
components overlaid (black line). The blueshifted and redshifted components
are detected at 3$\sigma$ and 4$\sigma$ significance,
respectively.}\label{fig:txs0828spec} \end{figure}

The integrated spectrum is shown in Fig.~\ref{fig:txs0828spec}. We find 2
compact components, labeled TXS0828+193 SW1 and SW2, respectively. They have
the same spatial position, but different blueshifts of $\Delta v_{\rm
  SW1}$$=$-200$\pm$40 km s$^{-1}$ and $\Delta v_{\rm SW2}$$=$-920$\pm$70 km s$^{-1}$
relative to the HzRG, respectively. The systemic velocity was estimated from
rest-frame optical integral-field spectroscopy of TXS0828+193
\citep[][]{nesvadba08b}. SW1 and SW2 have line widths FWHM$_{\rm
  SW1}$$=$310$\pm$270 km s$^{-1}$ and FWHM$_{\rm SW2}$$=$340$\pm$270 km s$^{-1}$,
respectively. Integrated fluxes are I$_{\rm CO,SW1}$$=$0.23$\pm$0.08 Jy km
s$^{-1}$ and I$_{\rm CO,SW2}$$=$0.24$\pm$0.06 Jy km s$^{-1}$, respectively, and
correspond to luminosities of ${\cal L}^{\prime}_{CO}$$=$9$\times$$10^9$ K km
s$^{-1}$ pc$^{2}$ per component. (We assumed ${\cal L}^{\prime}_{3-2}$$=$${\cal
  L}^{\prime}_{1-0}$ and $r_{32}$$=$1).

Molecular gas mass estimates of high-redshift galaxies 
depend on the assumption that the conversion factors from CO to
H$_2$ (``X factors'') established at low redshift will apply. The X-factor of
$\rm X_{\rm U}=0.8\ {\rm M_{\odot} / (\rm K\ \rm km\ \rm s^{-1}\ pc^{2})}$
appropriate for ULIRGs yields estimates $\sim 5\times$ lower than the Milky
Way X-factor \citep[][]{downes98}. With X$_{\rm U}$, the $\rm I_{\rm SW1}\sim
\rm S_{\rm SW2} \sim 0.25$ Jy km s$^{-1}$ per component correspond to $7\times
10^{9}$ M$_{\odot}$ in cold gas per component, or $1.4\times 10^{10}$ 
M$_{\odot}$ in total.

We will now discuss two possible scenarios for the nature of the CO emission
in the halo of TXS0828+193, namely, that it may be associated with a satellite
galaxy, or that it may be related to gas clouds or filaments within the
gas-rich halo of TXS0828+193.

\subsection{An extremely gas-rich satellite galaxy?}
\label{ssec:galaxy}
\begin{figure*}
\centering
\epsfig{figure=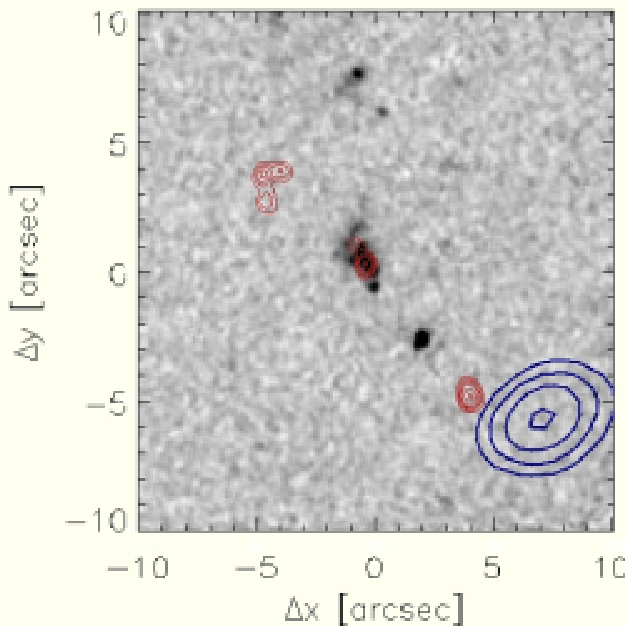,width=0.24\textwidth}
\epsfig{figure=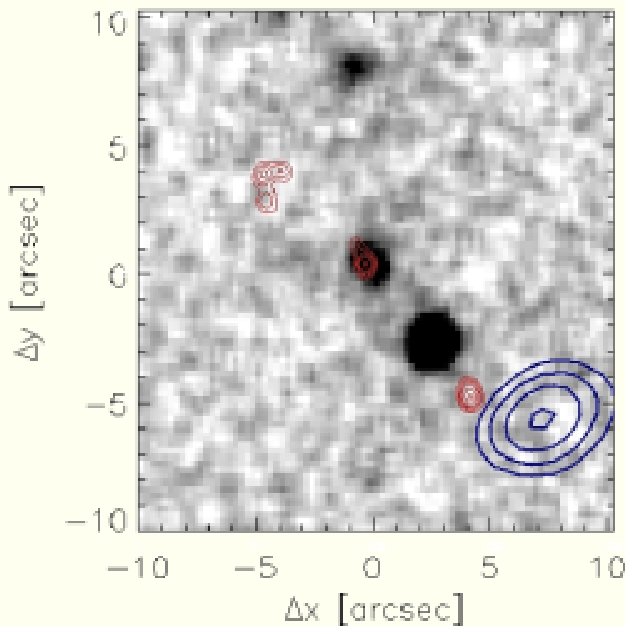,width=0.24\textwidth}
\epsfig{figure=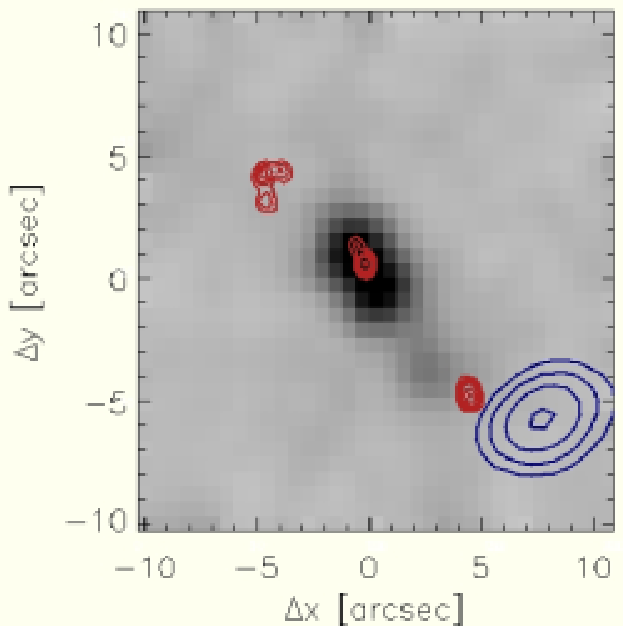,width=0.24\textwidth}
\epsfig{figure=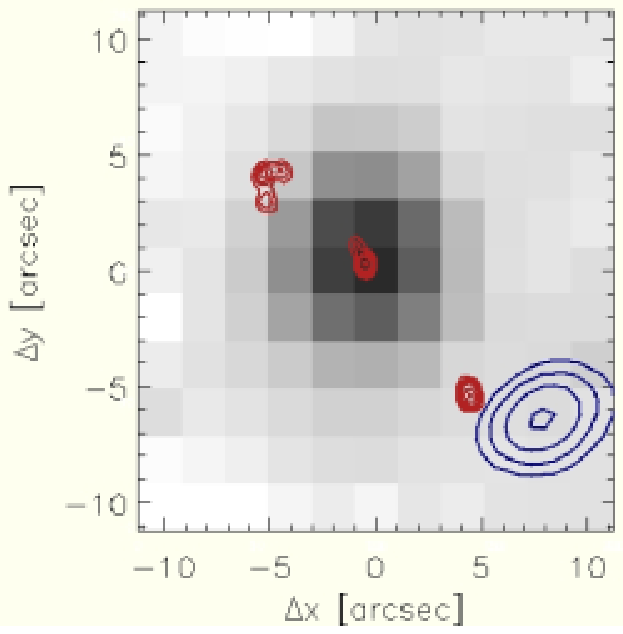,width=0.24\textwidth}

\caption{{\it (left to right:)} HST WFPC606W, Palomar K band, Spitzer IRAC
3.6$\mu$m and MIPS 24$\mu$m photometry of TXS0828+193. Thick blue contours
show the position of the SW1/SW2, thin red contours mark radio
jets. SW1/SW2 is undetected in all bands.}
\label{fig:txs0828_photometry}
\end{figure*}

We searched for the stellar continuum of putative galaxies associated with
SW1/2 in our set of images with rest-frame wavelengths between $\sim
1700$\AA\ and 7$\mu$m (\S\ref{sec:observation}). SW1/2 was not detected in any
of the data sets (Fig.~\ref{fig:txs0828_photometry}). This is in strong contrast to the companion of the z$=$3.8
HzRG 4C60.07, that was detected in all bands in a similar study of
\citet{ivison08}. We use the deep K-band photometry with a 3$\sigma$ limit of
K$_{\rm 3\sigma}$$=$23.7 mag in a 1\arcsec\ aperture and the population
synthesis models of \citet{bruzual03}, to place an upper limit on the stellar
mass. Continuous star-formation histories with ages between $5\times 10^7$ yrs
and $2\times 10^9$ yrs (implying a formation at z$\lesssim$10), and
extinctions A$_V =1-5$ mag correspond to a maximum of $\sim 3\times 10^9
M_{\odot}$ in stellar mass. This covers more than the range of extinctions
found for dusty submillimeter galaxies at z$\ge 2$ and low-redshift ULIRGs
typically associated with strong CO emission \citep[A$_V\lesssim 2$ mag][]{smail04,scoville00}. These extinctions are
derived from the integrated photometry of the galaxies, as appropriate for our
purposes. Extinctions along individual sightlines and into a starburst may be
significantly higher.

The r.m.s. of $\sim 100$ $\mu$Jy in our MIPS 24$\mu$m image is well below the
fluxes measured by \citet[][]{pope08} for submillimeter galaxies
at similar redshifts, which are in the range 200-500$\mu$Jy. This allows us to
set constraints on the star formation, because the filter covers the 6.2$\mu$m
PAH band, and more than half of the 7.7$\mu$m band at the redshift of
SW1/2. With the MIPS non-detection, it appears unlikely that SW1/2 
is forming stars at the prodigeous rates of several 100 M$_{\odot}$ typically
observed in SMGs. 

We will now estimate a dynamical mass for SW1/2. 
With a beam size of 40 kpc, we do not know whether the CO line emission may be
associated with one or with two 
galaxies. If SW1/2 represent a
double-horned profile of a single, roughly virialized, rotating galaxy (as
often assumed for SMGs), we can estimate a mass by setting M$_{dyn} = (v/\sin
i)^2\ R_{1kpc}\ / G$ with circular velocity $v=$350 km s$^{-1}$, inclination
$i$, radius $R_{1 kpc}$, and gravitational constant, $G$. The radius $R$ is
given in kpc. \citet[][]{tacconi08} use radii $\sim$ 2$-$5 kpc for their mass
estimates, which would imply M$_{dyn}\sim 0.6-1.5\times 10^{11}$ M$_{\odot}$
for an edge-on disc, and, perhaps more
realistically, 2-3$\times$ higher masses for a more average inclination. Thus,
SW1/2 would have a mass in the typical range of submillimeter galaxies or
powerful radio galaxies, which are K$\sim$20 mag or brighter
\citep[][]{smail04,debreuck02}. This would also significantly exceed the baryonic (gas and
stellar) mass of few $\times 10^{10}$ M$_{\odot}$.
If alternatively, we assume that SW1 and SW2 are associated with two different
galaxies in the halo of TXS0828+193 (a plausible assumption given the 40 kpc
covered by the beam), then, following \citet[][]{neri03}, we estimate a
dynamical mass of M$_{dyn}=4\times 10^9\ R_{1kpc}$ M$_{\odot}$ per galaxy for
a FWHM$=$300 km s$^{-1}$ line width of each component. Assuming a radius of a
few kpc, the dynamical mass estimate will be lower than the molecular gas mass
by factors of a few, except if we assume that both galaxies are seen within a
few degrees from being face-on, which does not appear very likely.
Each of these estimates relies on the assumption that
the gas is approximately virialized. This is a common assumption in CO
emission line studies at high redshift, but whether it is justified has yet to
be proven. In \S\ref{ssec:blob} we discuss 
a scenario where the virial assumption would not apply. Likewise
\citet{ivison08} raised doubts as to whether this assumption is always
justified in the context of high-redshift galaxies.

In these ``galaxy'' scenarios, we also need a mechanism to 
suppress star formation in the cold gas traced by the CO. If the CO line
emission arises from a disc with a few kpc in radius, then the observed gas
mass of $1.4\times 10^{10}$ M$_{\odot}$ corresponds to a surface mass density of
few $\times$ 1000 M$_{\odot}$ pc$^{-1}$. Following the Schmidt-Kennicutt
relation \citep[][]{kennicutt98} between gas surface density and star
formation intensity, SFI, we expect SFI $=$ few $\times$ 10 M$_{\odot}$
yr$^{-1}$ kpc$^{-2}$. Averaging over the size of the disc, this corresponds
to a total star-formation rate of several 100 M$_{\odot}$ yr$^{-1}$. This is
in the typical range of submillimeter galaxies, but in contradiction
with our non-detection at 24$\mu$m. Likewise, with star-formation rates of
few$\times 100$ M$_{\odot}$ a stellar mass of few$\times 10^9 M_{\odot}$ would
be built in few$\times 10^7$ yrs, so that SW1/2 would have to be in a very
special, very young stage of the starburst if it was a
galaxy.

\citet{papadopoulos08} recently found luminous, but excited CO line
emission in a nearby radio galaxy, 3C293, which does not seem associated with
a starburst. The same is suggested by Spitzer observations of a small number
of nearby galaxies with strongly enhanced, mid-infrared H$_2$ line emission
\citep[e.g.,][]{ogle07}. \citet{guillard08} argue that this gas may be heated
through the dissipation of kinetic energy. In these cases, it is likely that
the energy was injected by an external mechanism (a merger or AGN). We will in
the following propose a somewhat related scenario, where the nearby radio lobe
may have induced the collapse of gas within the halo of TXS0828+193.

\subsection{Cold gas in the halo?}
\label{ssec:blob}

\begin{figure}
\centering
\epsfig{figure=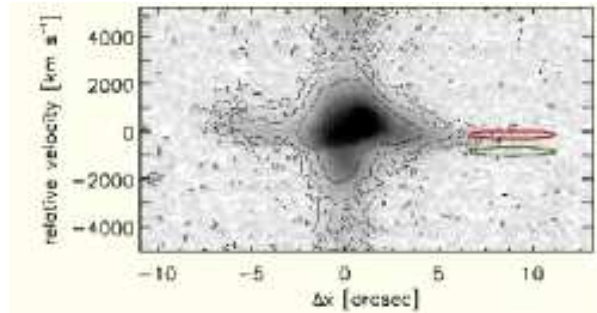,width=0.45\textwidth}
\caption{Ly$\alpha$ spectrum of TXS0828+193, graciously provided by
M. Villar-Martin with the CO(3--2) components SW1 and SW2 shown as red and
green ellipses, respectively. Positions and sizes correspond to the distance
from the HzRG and beam size, and to the relative redshift and FWHM line width
of the two components, respectively.}
\label{fig:txs0828_spectrum}
\end{figure}

TXS0828+193 was the first HzRG where an outer halo was found \citep{villar02},
which extends beyond the turbulent, luminous inner emission line region that
is likely powered by energy released from the powerful AGN
\citep[e.g.,][]{villar99,nesvadba08b}. Diffuse gas in the outer halo is
fainter, and has more moderate velocity gradients and line widths, which may
indicate rotation in the potential of the underlying dark-matter halo
\citep{villar03,villar06}, or perhaps gas infall \citep{humphrey07}.
\citet{villar02} find CIV emission in the outer halo of TXS0828+193 and
suspect that metallicities may be up to nearly solar, perhaps representing gas
that has previously been driven out from the central galaxy.

Fig.~\ref{fig:txs0828_spectrum} shows the relative velocity and radial
distance of SW1 and SW2 from TXS0828+193 relative to the two-dimensional
Ly$\alpha$ longslit spectrum of \citet{villar02}. The slit of the Ly$\alpha$
spectrum falls near the CO position, but since the slit width, 1\arcsec, is
significantly smaller than the beam of 5\arcsec, we cannot infer whether it
covers the position of the CO emitter. Ly$\alpha$ emission is traced almost
out to the 24 kpc radial distance of SW1/2 from the HzRG.

SW1 and SW2 are blueshifted relative to the radio galaxy, and are also on the
side of the radio galaxy where the diffuse Ly$\alpha$ halo is blueshifted (the
luminous inner halo is redshifted). The kinematic properties of SW1
(FWHM$=$310$\pm$270 km s$^{-1}$, z$=$2.5761$\pm$0.0008) are very similar to
those of the diffuse ionized gas (FWHM=590$\pm$60 km s$^{-1}$,
z$=$2.5741$\pm$0.0008), with a relative blueshift of SW1 relative to the
Ly$\alpha$ emission of $170\pm60$ km s$^{-1}$. SW2 (FWHM$=340\pm 270$,
z$=$2.5678$\pm$0.0007) is blueshifted by 890 km s$^{-1}$ relative to the
diffuse Ly$\alpha$ emission, but has a similar line width within large
uncertainties. Given the low stellar mass possibly associated with the CO line
emission (\S~\ref{ssec:galaxy}), this may suggest that the CO emission
originates from clouds or filaments in the diffuse halo gas of TXS0828+193,
which has metallicities of up to about solar \citep{villar02}.
\citet{debreuck03} detected CO emission at the
redshift of a Ly$\alpha$ absorber near the z$=$3.1 HzRG B2 2330$+$3927, and
proposed a similar scenario, but did not have the spatial resolution to
directly measure positional offsets between the radio galaxy and CO emission.

Several mechanisms may plausibly influence the halo gas including
minor or major mergers, or powerful outflows from starbursts and AGN. Each of
these mechanisms may sweep up and accelerate halo gas over timescales
of a few $\times 10^8$ yrs similar to those suggested by the relative velocity
of the CO and distance to the radio galaxy. The close proximity of the
emitters to the radio hot spot and alignment with the jet axis
is however suspicious. The emitters are very close to the jet axis, and within
a projected area of $\sim$300 kpc$^2$ from the hot spot, whereas our data have a
half-power beam width covering a total of 135,000 kpc$^2$. If this is not due
to mere projection effects, then weak shocks produced by the expanding radio
source may play a role in compressing and exciting the gas (the radio hot spot
is only about 2.5\arcsec\ or 20 kpc away, and we may not detect the
low surface brightness radio plasma). In turn, the interaction with dense gas
may be enhancing or even triggering the radio hot spot.

Extended cold molecular gas is found in some massive cooling-flow clusters at
low redshift \citep[e.g.,][]{edge01,salome04}, where molecular gas forms along
the edges of X-ray cavities inflated by the radio jet and outside the volume
filled by the radio plasma. Similarly, CO line emission in the halos of HzRGs
may trace the edges of cavities inflated by the radio lobes. Within the large
uncertainties, the $\sim$300 km s$^{-1}$ FWHM of SW1/2 are not very different
from the line widths in the diffuse CO in the Perseus cluster
\citep[][]{salome08a}.

We can use the observed surface brightness of the faint Ly$\alpha$ emission in
the halo of TXS0828+193 and the observed CO luminosity to investigate whether
TXS0828+193 falls near the correlation between the luminosity of the molecular
and ionized gas found in local cooling-flow clusters \citep{edge01}. Adopting
a flux conversion ${\cal L}_{Ly\alpha} = 13\times {\cal L}_{H\alpha}$ between
Ly$\alpha$ and H$\alpha$ luminosity (where we neglect extinction), we find a
strict upper limit on the Ly$\alpha$ luminosity of ${\cal L}_{Ly\alpha} =
1.4\times 10^{42}$ erg s$^{-1}$ cm$^{-2}$ within the $20\square$\arcsec\ area
of the beam. Translating the CO gas mass of \citet{edge01} into a CO
luminosity, we find that the halo of TXS0828+193 falls only factors of a few
below the expected value found in local cooling-flow clusters. Allowing for
different physical conditions, gas distributions, and AGN properties (HzRGs
host powerful AGN), we may well be seeing a fundamentally similar
phenomenon.

However, the brightness temperatures and spatial distribution of the extended
CO emission in low-redshift clusters are significantly different. Low
brightness temperatures suggest low gas filling factors, and much of the gas
is concentrated towards the central radio galaxy. This may arise from
different properties of the diffuse cluster gas at high and low redshift,
which is cold for HzRGs \citep[with embedded filaments or clouds of few
$\times 10^9$ M$_{\odot}$ and more in dense neutral gas][]{vanojik97} and hot
and rarefied in low-redshift clusters. In fact, in HzRGs at z$\sim$ 2-3 we
may be witnessing the processes that led to the rapid heating and entropy
enhancement of cluster gas through AGN feedback, which seem necessary to
explain the temperature profiles of massive X-ray clusters at low redshift
\citep{nath02,mccarthy08}. Interestingly, Ly$\alpha$ absorbers are only found
in the halos of HzRGs with relatively small, and likely rather young, radio
sources. This suggests that large amounts of dense gas (and perhaps dust) may
be present in the 
halo at radii that are not yet affected by mechanical heating from the radio
source. The approaching radio jet of TXS0828+193 may have contributed to
triggering the collapse and forming SW1/2. Similar processes may ultimately
lead to positive AGN feedback and jet-triggered star formation in
some cases, if the cold molecular gas will relax and form stars over
sufficiently short time scales.

\section{Summary and Conclusions}
We discuss the nature of luminous CO(3--2) line emission in the halo of the
radio galaxy TXS0828+193 at z$=$2.6. The CO emission resembles that
of submillimeter galaxies, but we do not detect continuum emission from SW1/2,
including 24$\mu$m MIPS imaging, which covers the PAH bands at z$=$2.6. For a
gas disc in a galaxy we would expect strong star formation if the
Schmidt-Kennicutt law roughly applies.

Alternatively, SW1/2 may represent a cloud or filament in the halo, maybe
related to neutral, dense Ly$\alpha$ absorbers observed near some HzRGs. The
approaching radio jet of TXS0828+193 may have contributed to triggering the
collapse and exciting the gas. This is somewhat in analogy with diffuse CO
emission in low-redshift clusters, but the ambient gas properties will likely
be very different at z$=$2.6.  In either case, SW1/2 does not appear to be an
'ordinary' high-redshift CO emitter, and further observations are necessary to
differentiate between the two scenarios. 
This suggests that CO
observations of the high-redshift Universe with the refurbished PdBI and soon
with ALMA, will reveal a rich, and multi-faceted picture of the early
Universe.

\section*{Acknowledgments}
We would like to thank the staff at IRAM for carrying out the observations and
for 
hospitality during the data reduction. We also thank M. Villar-Martin
and C. Carilli for valuable discussion and for generously sharing their
data. NPHN thanks P.~Salom\'e, G. Bicknell, and M. Krause for interesting
discussions. We thank the referee for comments which helped improve the
paper. NPHN acknowledges financial support through a fellowship of the
Centre National d'Etudes Spatiales (CNES) and through a Marie Curie Fellowship
of the European Commission. IRAM is funded by
the Centre National de Recherche Scientifique, the Max-Planck Gesellschaft and
the Instituto Geografico Nacional.

\bibliography{nesvadba}

\end{document}